\documentclass[sigconf]{acmart}%
\usepackage[utf8]{inputenc} % allow utf-8 input
\usepackage[T1]{fontenc}    % use 8-bit T1 fonts
\usepackage{hyperref}       % hyperlinks
\usepackage{amsfonts}       % blackboard math symbols
\usepackage[ruled]{algorithm2e}
\usepackage{graphicx}
\usepackage{multirow}
\usepackage{float}
\usepackage{amsmath} 
\usepackage{amsthm}
\usepackage{graphics}
\usepackage{subfigure}
\AtBeginDocument{%
  }

\setcopyright{acmcopyright}
\copyrightyear{2023} 
\acmYear{2023} 
\setcopyright{acmlicensed}\acmConference[MM '23]{Proceedings of the 31st ACM International Conference on Multimedia}{October 29-November 3, 2023}{Ottawa, ON, Canada}
\acmBooktitle{Proceedings of the 31st ACM International Conference on Multimedia (MM '23), October 29-November 3, 2023, Ottawa, ON, Canada}
\acmPrice{15.00}
\acmDOI{10.1145/3581783.3612591}
\acmISBN{979-8-4007-0108-5/23/10}

\begin{document}

\title{Pareto Invariant Representation Learning for Multimedia Recommendation}

\author{Shanshan Huang}
\authornote{Both authors contributed equally to this research.}
%\orcid{0000-0001-7893-3861}
\affiliation{%
 \institution{%School of Big Data \& Software Engineering\\ 
 Chongqing University}
 % \streetaddress{P.O. Box 1212}
 \city{Chongqing}
 \country{China}
 \postcode{400044}
 }
 \email{shanshanhuang@cqu.edu.cn}
 
\author{Haoxuan Li}
\authornotemark[1]
\affiliation{%
 \institution{Peking University}
 % \streetaddress{P.O. Box 1212}
 \city{Beijing}
 \country{China}
 \postcode{100091}
}
\email{hxli@stu.pku.edu.cn}

\author{Qingsong Li}
\affiliation{%
 \institution{%School of Big Data \& Software Engineering\\ 
 Chongqing University}
 % \streetaddress{1 Th{\o}rv{\"a}ld Circle}
 \city{Chongqing}
 \country{China}
 \postcode{400044}
 }
 \email{liqingsong@stu.cqu.edu.cn}

\author{Chunyuan Zheng}
\affiliation{%
 \institution{University of California, San Diego}
 % \streetaddress{P.O. Box 1212}
 \city{San Diego}
 % % \state{Ohio}
 \country{USA}
 % \postcode{XX}
 }
 \email{czheng@ucsd.edu}

\author{Li Liu}
\authornote{Corresponding author.}
\affiliation{%
 \institution{%School of Big Data \& Software Engineering\\ 
 Chongqing University}
 % \streetaddress{1 Th{\o}rv{\"a}ld Circle}
 \city{Chongqing}
 \country{China}
 \postcode{400044}
 }
 \email{dcsliuli@cqu.edu.cn}
%

% \renewcommand{\shortauthors}{Huang et al.}
\renewcommand{\shortauthors}{Shanshan Huang, Haoxuan Li, Qingsong Li, Chunyuan Zheng, \& Li Liu}

\begin{abstract}
Multimedia recommendation involves personalized ranking tasks, where multimedia content is usually represented using a generic encoder. However, these generic representations introduce spurious correlations that fail to reveal users' true preferences. Existing works attempt to alleviate this problem by learning invariant representations, but overlook the balance between independent and identically distributed (IID) and out-of-distribution (OOD) generalization. In this paper, we propose a framework called \underline{Pa}reto \underline{Inv}ariant \underline{R}epresentation \underline{L}earning (PaInvRL) to mitigate the impact of spurious correlations from an IID-OOD multi-objective optimization perspective, by learning invariant representations (intrinsic factors that attract user attention) and variant representations (other factors) simultaneously. Specifically, PaInvRL includes three iteratively executed modules: (i) heterogeneous identification module, which identifies the heterogeneous environments to reflect distributional shifts for user-item interactions; (ii) invariant mask generation module, which learns invariant masks based on the Pareto-optimal solutions that minimize the adaptive weighted Invariant Risk Minimization (IRM) and Empirical Risk (ERM) losses; (iii) convert module, which generates both variant representations and item-invariant representations for training a multi-modal recommendation model that mitigates spurious correlations and balances the generalization performance within and cross the environmental distributions. We compare the proposed PaInvRL with state-of-the-art recommendation models on three public multimedia recommendation datasets (Movielens, Tiktok, and Kwai), and the experimental results validate the effectiveness of PaInvRL for both within- and cross-environmental learning.
\end{abstract}





% \received{20 February 2023}

\maketitle

\section{Introduction}
With the rapid development of the internet, multimedia recommendation systems have become indispensable tools to help users find their interesting items, and have been widely used in many online applications, such as e-commerce platforms, social media, and instant video platforms. For multimedia recommendation, item content includes multiple modalities, including visual, acoustic, and textual representations. These multi-modal data may reflect user preferences at the fine-grained modality level. The core of multimedia recommendation is to use the historical interactions between users and items and the auxiliary multi-modal item representations to improve recommendation performance.

Collaborative filtering (CF) serves as the foundation of personalized recommendation systems, which leverages historical user-item interactions to learn user and item representations and provides recommendations based on these representations \cite{su2009survey, schafer2007collaborative}. 
Extending to multimedia tasks, previous studies, e.g., VBPR \cite{he2016vbpr}, DeepStyle \cite{liu2017deepstyle}, incorporate multi-modal contents as side information in addition to id embeddings of items to learn the user preference. However, these methods have limited expressiveness as they neglect high-order user-item semantic relations \cite{zhou2023comprehensive}. Inspired by the recent advances in graph neural networks, recent studies \cite{wu2019session, wang2019neural, sun2020multi, guo2021deep, liu2022elimrec, lei2023learning} take advantage of powerful graph convolution networks (GCNs) to model user-item relationships as bipartite graphs to improve the performance of CF-based recommendation systems. Further, many researchers have also attempted to apply GCNs to incorporate modality information into the message passing for inferring user and item representations, such as MMGCN \cite{wei2019mmgcn}, GRCN \cite{wei2020graph}, LATTICE \cite{zhang2021mining}, MICRO \cite{zhang2021mining} and HCGCN \cite{mu2022learning}.%cai2022adaptive, 

Despite achieving promising performance, previous approaches often use encoder architectures designed for general content understanding tasks \cite{HUANG2022105006} (including image classification, object recognition, image colorization, and text classification, etc.), e.g., pre-trained VGG19 \cite{simonyan2014very}, ResNet50 \cite{he2016deep}, VILBERT \cite{lu2019vilbert}, and sentence-transformer \cite{reimers2019sentence}, to encode multimedia content. The use of these generic encoders may introduce spurious correlations (i.e., some learned representations may affect the recommendation results, but are irrelevant to user's true preferences from a causal perspective), making it difficult for recommendation models to capture user's true preferences and provide accurate recommendations. To alleviate this issue, existing studies mainly rely on preference-aware representations \cite{yu2018aesthetic,liu2019user,shen2020enhancing}, which were extracted with specifically designed multimedia models for specific recommendation tasks. Therefore, the existing methods face the limitation of domain-specific analysis and design, and thus can hardly be generalized. 

To address this issue, a recent research work, named InvRL \cite{du2022invariant}, introduced invariant risk minimization (IRM) to multimedia recommendation, by learning invariant item representations to alleviate the impact of spurious correlations. Although experimentally promising, it is widely known that there is a conflict between independent and identical distributed (IID) tasks (where the source and target environments are similar) and out-of-distribution (OOD) tasks (where there is a significant difference between the source and target environments), which may lead to significant degradation of model performance on IID tasks. We verify empirically that the superiority of InvRL is only guaranteed in OOD tasks, whereas empirical risk minimization (ERM) typically outperforms in IID tasks, which motivates us to
% it is known that IRM conflicts with empirical risk minimization(ERM) and therefore may lead to significant degradation of ERM performance. 
% Additionally, the performance of IRM is guaranteed only when there is a significant difference between the source and target environments(i.e., out-of-distribution problem, OOD), and the performance of empirical risk minimization (ERM) is typically better when the source and target environments are similar (i.e., independent identical distribution problem, IID). 
balance this conflict between IID and OOD. Specifically, in this paper, we formalize the IID-OOD task as a multi-objective optimization problem \cite{zhang2021survey} and adaptively weight the ERM and IRM losses via a gradient approach to obtain the Pareto optimal solution. We theoretically prove that our solution cannot be dominated by other solutions, i.e., there does not exist any solution that performs better compared to our solution on both tasks at the same time. Specifically, we divide the raw multimedia representations into two parts: variant and invariant representations, where the variant representations account for spurious correlations while the invariant representations reflect the user's true preferences.

% To obtain invariant representations, learning invariant masks is the key, so we first follow HRM \cite{liu2021heterogeneous} to automatically classify heterogeneous environments by historical user-item interactions, and then use adaptively weighted IRM loss and ERM loss to train the recommendation model to learn invariant masks from the partitioned environments and obtain the final invariant representations. Finally, the learned invariant representations are used to promote the final recommendation model. To verify the effectiveness of the proposed method, we conducted a series of experiments on three public datasets. The experimental results show that our method achieves state-of-art multimedia recommendation performance.
The main contributions of this paper are summarized as follows.\vspace{-12pt}
\begin{itemize}
	\item We first formalize the IID-OOD task as a multi-objective optimization problem and adaptively weight the ERM and IRM losses using a gradient-based representation learning approach to obtain the Pareto optimal solution, i.e., there does not exist any solution that outperforms compared to our solution on both IID and OOD tasks.
	\item  We propose a new multimedia recommendation framework, called PaInvRL, that aims to obtain a Pareto solution between IID and OOD tasks via a gradient-based updating method, where the gradient is shown to be either 0 when there are no other solution in its neighborhood can have lower values in both ERM and IRM losses, or the gradient givens a descent direction that improves both IID and OOD generalization by reducing ERM and IRM losses simultaneously.
	\item We instantiate the framework over UltraGCN and conduct extensive experiments over three public datasets, verifying the rationality and effectiveness of PaInvRL.
\end{itemize}

\section{Related work}
\vspace{-0.7pt}
\subsection{Multimedia Recommendation}
\vspace{-0.7pt}
The multi-modal recommendation system aims to learn informative representations of users and items by leveraging multi-modal representations. Many efforts \cite{ma2019disentangled, ma2019learning, mao2021ultragcn, wei2021hierarchical} have been devoted to enhancing recommendation systems by incorporating multimedia content. VBPR \cite{he2016vbpr} is the first model that considers introducing visual representations into the recommendation system by concatenating visual embeddings with id embeddings as the item representations. DVBPR \cite{kang2017visually} attempts to jointly train the image representations as well as the parameters in a recommendation model. In recent years, graph neural networks have been demonstrated as powerful solutions for multimedia recommendation by capturing high-order dependent structures among users and items. For example, MMGCN \cite{wei2019mmgcn} constructs a modal-specific graph and conducts graph convolution operations, to capture the modal-specific user preference and distills the item representations simultaneously. MGAT \cite{tao2020mgat} based on the MMGCN framework utilizes the original GCN to do aggregation and the same way to combine the aggregated result. To manage the information transmission for each modality, it added a new gated attention mechanism. DualGNN \cite{wang2021dualgnn} also introduces a model preference learning module and draws the user’s attention to various modalities. InvRL \cite{du2022invariant} introduces IRM to learn invariant item representations, which reduces the impact of spurious correlations and improves the recommendation performance of multi-modal recommendation models. DRAGON \cite{zhou2023enhancing} learns dual representations of users and items by constructing homogeneous graphs to enhance the relationship between the two parties, enabling multi-modal recommendations. Different from these works, for robust multi-modal user preference learning, this paper proposes a new framework for invariant representation learning, which first views the IID-OOD task in the multi-modal recommendation as a multi-objective optimization problem, and then adaptively weights the IRM and ERM losses and uses gradient-based methods to seek Pareto optimal solutions for learning invariant representations.

\vspace{-7.5pt}
\subsection{Invariant Representation Learning}
\vspace{-0.7pt}
Invariant representation learning aims to learn the essential representations of data, and improve the generalization ability and robustness of models. Recently, some studies have been conducted, among which IRM \cite{arjovsky2019invariant} is an earlier method proposed based on invariant principle \cite{peters2016causal}, which aims to learn representations with invariance in different environments. Several works \cite{ahuja2020invariant,krueger2021out, zhou2022sparse, lin2022zin} further develop several variants of IRM by introducing game theory, regret minimization, variance penalization, etc., and \cite{xu2021learning, xu2022regulatory} try to learn invariant representations by coupled adversarial neural networks. Other approaches \cite{liu2021kernelized, zhang2022towards} attempt to learn invariant representations without providing explicit environment indicators. Liu et al. \cite{liu2021heterogeneous} proposed the HRM to achieve joint learning of latent heterogeneity and invariant relationships in the data, resulting in stable predictions despite distributional shifts. Furthermore, they extended HRM to the representation level using kernel tricks \cite{liu2021kernelized}. 

An alternative class of methods for learning invariant representations are causality-based approaches with debiased loss~\cite{wang2020information,Chen-etal2020,liu2021mitigating,Wu-etal2022,wang2023causal,Li-etal2023KDD,bonner2018causal,Chen-etal2021,liu2020general,Wang-etal2021,Balance2023,zeyukdd23,li-noharm-2023}, such as outcome regression methods~\cite{Hansotia-Rukstales2001, imbens2015causal,Steck2010, marlin2012collaborative,yuan2020unbiased}, propensity-based weighting methods~\cite{li-propensity-2023,Horvitz-Thompson1952, Nie-Wager2021, Hernan-Robins2020,Schnabel-Swaminathan2016,yuan2019improving,li-propensity-2023}, doubly robust learning methods~\cite{Robins-Rotnitzky-Zhao1994, Kennedy-2020,Wang-Zhang-Sun-Qi2019,saito2020doubly,MRDR,Dai-etal2022,TDR,li2022-SDR,wang2022escm2}, multiple robust learning method~\cite{MR2023}, and representation learning methods~\cite{chipman2010bart, wager2018estimation,johansson2016learning,shalit2017estimating,schwab2018perfect,kunzel2019metalearners,shi2019adapting,zhong2022descn}.  However, these previous approaches failed in obtaining Pareto solutions between IID and OOD tasks \cite{sener2018multi,lin2019Pareto}. In this paper, we aim to learn representations corresponding to the Pareto solutions between within- and cross-environmental learning to improve the model's generalization performance in multimedia recommendations.

\begin{figure*}[t]
	\centering
	\includegraphics[width=0.9\linewidth]{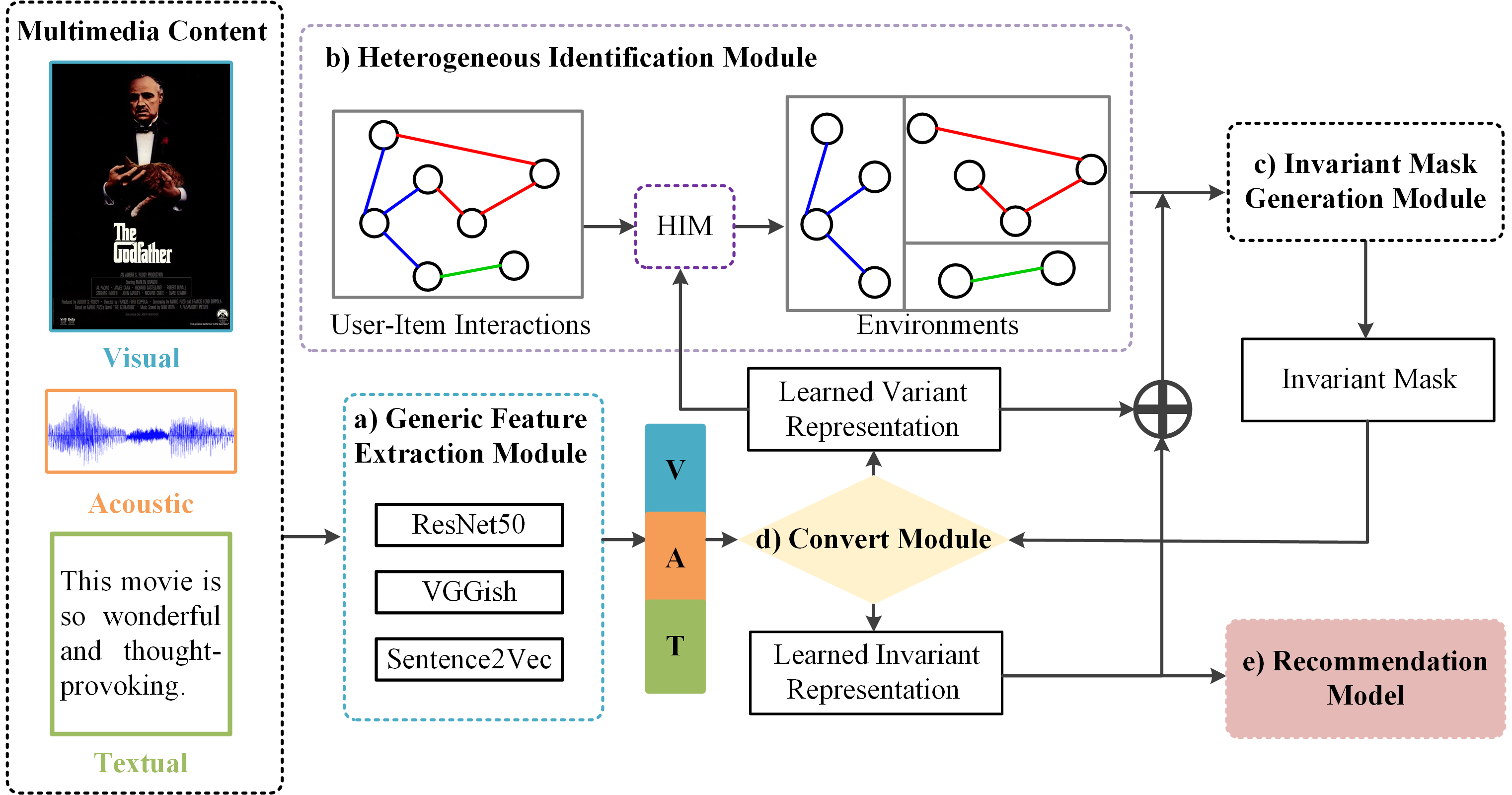}
        \vspace{-9pt}
	\caption{The framework of PaInvRL, where V, A, and T denote the extracted visual representations, acoustic representations, and textual representations, respectively. The symbol $\oplus$ represents the operation of weighted summation.}
    \vspace{-9pt}
	\label{fig:framework} 
\end{figure*}
\section{Methodology}
\subsection{Preliminaries}
Considering a multimedia recommendation system, we denote the set of users and items as $\mathcal{U}$ and $\mathcal{I}$, respectively. For each user-item pairs $(u, i) \in \mathcal{U}\times \mathcal{I}$, denote $r_{u, i} = 1$ if user $u$ make a positive feedback on item $i$, and $r_{u, i}=0$ otherwise. In addition to user-item interactions, we also have access to multi-modal representations that provide content information about items. We represent the modality representation of item $i$ as $\mathbf{f}_{r,i} \in \mathbb{R}^{d_r}$, where $d_r$ is the dimension of the modality representation, $r \in R=\{V, T, A\}$ denotes the modality, and $R$ is the set of all modalities. In this paper, $R$ includes visual (V), textual (T), and acoustic (A) modalities, let ${\mathbf{f}_i} = concat({\mathbf{f}_{V,i}}, {\mathbf{f}_{T,i}}, {\mathbf{f}_{A,i}})\in \mathbb{R}^{d}$, where $d=d_V+d_T+d_A$ and $concat(\cdot)$ indicates the concatenation operation. The multi-modal recommendation aims to learn a model $\Gamma(u, i, \mathbf{f}_i| \Theta)$ parameterized by $\Theta$ to predict users’ true preferences, which can be formalized as
\begin{equation}
	% \mathop {\arg \min }\limits_\Theta  \mathcal{L}(T(u,i,f_i)|{\mathcal R^{tr}})
\arg \min_{\Theta}  \mathcal{L}(\Gamma(u,i,\mathbf{f}_i | \Theta)|{\mathcal R^{tr}}),
	\label{eq:problemF}
\end{equation}
where $\mathcal{L}(\cdot)$ denotes the recommendation loss, and $\mathcal R^{tr}$ denotes the training set, with both positive samples $\mathcal R^+=\{(u, i): r_{u, i} = 1\}$ and negative samples $\mathcal R^-=\{(u, i): r_{u, i} = 0\}$.

\subsection{Model Overview}
We now present the proposed PaInvRL model, the architecture of which is illustrated in Figure \ref{fig:framework}. There are four components in the framework: (1) the generic feature extraction network that is used to extract multi-modal representations, including visual, acoustic, and visual representations; (2) the heterogeneous identification module (HIM) that is designed to partition the input historical user-item dataset interaction into multiple heterogeneous environments for invariant representation learning, each reflecting a spurious correlation in user-item interactions; (3) the invariant mask generation module and (4) the convert module work together to select representations that have stable and invariant relationships across environments. Specifically, the generic feature extraction module adopts a pre-trained model and is not the focus of this paper, and we therefore provide only a brief introduction to this module in Section \ref{section: 4}. The HIM and the invariant mask generation module promote each other: on one hand, the invariant mask generation module uses the heterogeneous environment identified by HIM to learn the invariant mask $\mathbf{m}$, which leads to the corresponding invariant representations $\mathbf{\Phi}_i$ and variant representations ${\mathbf{\Psi} _i}$ using the learned invariant mask; on the other hand, the variant representations are utilized to enhance the training of HIM. The convert module divides the raw multimedia representations into invariant representations and variant representations. Finally, we use the learned invariant representations to learn the final multi-modal recommendation model with both promising IID and OOD generalization. 

Different from InvRL \cite{du2022invariant} that utilizes the invariant mask generation module to generate invariant masks used to generate the corresponding invariant representations with superior performance under the OOD task, we propose to generate the invariant mask corresponding to a Pareto solution between IID and OOD tasks via a gradient-based updating method. The proposed invariant mask update gradient is either 0 when no neighboring solution can offer lower values in both ERM and IRM losses, or it provides a descent direction enhancing IID and OOD generalization through simultaneous reduction of ERM and IRM losses.

%Note that the generic feature extraction module is not the focus of this paper and thus has not been elaborated on. 
\subsection{Heterogeneous Environment Identification}
Heterogeneous identification module (HIM), which takes in the historical user-item interactions and outputs an environment set $\mathcal E$ for invariant mask generation \cite{du2022invariant}. This module comprises two phases: an environment learning phase and a user-item interaction partitioning phase. Specifically, in the environment learning phase, we learn different environments $e \in \mathcal E$ by training a recommendation model ${\Gamma_{(e)}(u, i, \mathbf{\Psi} _i}|\Theta _e)$ for each environment $e\in\mathcal{E}$, where $\Theta _e$ denotes the parameters of the recommendation model $\Gamma_{(e)}$, and can be optimized by
\begin{equation}
	\arg \min _{\Theta _e} \mathcal{L}({\Gamma_{(e)}(u, i, \mathbf{\Psi} _i}|\Theta _e)|\mathcal R_e^{tr}),
	\label{eq:HIM}
\end{equation}
where the variant representations $\mathbf{\Psi} _i$ are obtained by initializing the invariant mask $\mathbf{m}$ by 0.5. 
% The details of this operation are described in Section \ref{section: 3.5}. Each environment $e$ is obtained via an independent recommendation model $\Gamma_{(e)}$.
We employ UltraGCN \cite{mao2021ultragcn} as the recommendation model and drive the representations through a graph-based loss function to encode the user-item graph by 
\begin{equation}
	\mathcal{L} = {\mathcal{L}_O} + \eta {\mathcal{L}_U} + \kappa {\mathcal{L}_I},
	\label{eq:ultragcnloss}
\end{equation}
where  $\mathcal{L}_O$ is used as the main optimization objective of the recommendation model $\Gamma(u, i, {\mathbf{\Psi} _i})$, and $\mathcal{L}_U$ and $\mathcal{L}_I$ are used as constraints to learn better user-item graphs, and item-item graphs, respectively. $\eta$ and $\kappa$ are used as weights of $\mathcal{L}_U$ and $\mathcal{L}_I$ to adjust the relative importance of user-item and item-item relationships. Following \cite{mao2021ultragcn}, we choose the binary cross entropy loss to calculate $\mathcal{L}_O$ by 
\begin{equation}
	{\mathcal{L}_O} =  - \sum\limits_{(u,i) \in {\mathcal R^ + }} {\log ( {\sigma ( {\Gamma( {u,i,{\mathbf{\Psi} _i}} )} )} )} 
 - \sum\limits_{(u,j) \in {\mathcal R^ - }} {\log( {\sigma ( { - \Gamma( {u,j,{\mathbf{\Psi} _j}} )} )} )}, 
	\label{eq:LO}
\end{equation}
where $\sigma$ is the sigmoid function. $\mathcal{L}_U$ is derived by negative log-likelihood, as 
\begin{equation}
\begin{aligned}
	%\begin{array}{c}
		{\mathcal{L}_U} =  &- \sum\limits_{(u,i) \in {\mathcal R^ + }} {{\nu _{u,i}}\log ( {\sigma ( {\Gamma( {u,i,{\mathbf{\Psi} _i}} )} )} )} \\
		&- \sum\limits_{(u,j) \in {\mathcal R^ - }} {\nu _{u,j}} {\log ( {\sigma ( { - \Gamma( {u,j,{\mathbf{\Psi} _j}} )} )} )}, 
	%\end{array}
	\label{eq:LU}
\end{aligned}
\end{equation}
where $\nu _{u,i}$ and $\nu _{u,j}$ can be derived from the user-item graph by
\begin{equation}
	{\nu _{u,i}} = \frac{1}{{{d_u}}}\sqrt {\frac{{{d_u} + 1}}{{{d_i} + 1}}}, 
\end{equation}
where $d_u$ and $d_i$ denote the degrees for the corresponding nodes. The term $\mathcal{L}_I$ induced from item-item graph can be calculated by
\begin{equation}
    \mathcal{L}_I = -\sum\limits_{(u,i) \in {\mathcal R}^+} \sum\limits_{j \in {S(i)}} s_{i,j} \log( \sigma(\Gamma(u,j,\mathbf{\Psi} _j)),
\end{equation}
where $S(i)$ include $K$ weighted positive sample pairs $(u,k)$ corresponding to each positive sample pair $(u, i)$, which are selected from the weighted adjacency matrix of the item-item co-occurrence graph $G$ according to the similarity score $s_{i, j}$. We calculate $s_{i, j}$ by 
% Specifically, we first select $K$ weighted positive sample pairs $(u,k)$ for each positive sample pair $(u, i)$ from $G$ according to the similarity score $s_{i, k}$. $G$ is the weighted adjacency matrix of the item-item co-occurrence graph, and similarity score $s_{i, k}$ can be calculated by %$(u,j)$
\begin{equation}
	{s_{i,j}} = \frac{{{G_{i,j}}}}{{{g_i} - {G_{i,i}}}}\sqrt {\frac{{{g_i}}}{{{g_j}}}}, \quad {g_i} = \sum\limits_{k=1}^{K} {{G_{i,k}}}, 
\end{equation}
where $G_{i,j}$ represent the number of co-occurrences of item $i$ and item $j$, and $g_i$ and $g_j$ denote the degrees of item $i$ and item $j$ in $G$.

In the user-item interaction partitioning phase, we use the trained recommendation model to partition the user-item interaction into the corresponding environments by
\begin{equation}
	e \left( {u,i} \right) = \arg \max_{e \in \mathcal E} \Gamma_{(e)}(u, i, {\mathbf{\Psi} _i}| \Theta_e ).
	\label{eq:IP}
\end{equation}	
The obtained results $\{\mathcal R_{(e)}|e\in \mathcal E\}$ are used in the training of the following invariant mask generation module.

\subsection{Invariant Mask Generation}
Here, we introduce our invariant mask generation module, which takes multiple environments training data $\{\mathcal R_{(e)}|e\in \mathcal E\}$ as input, and outputs the corresponding invariant mask $\mathbf{m}$. As mentioned above, we learn the invariant mask generation module together with the convert module to generate invariant and variant representations across environments. Following InvRL \cite{du2022invariant}, we approximate $\mathbf{m} = (m_1, m_2, m_3, . . . , m_d)^T$ using clipped Gaussian random variable parameterized by $\mathbf{\mu} = (\mu_1, \mu_2, \mu_3, . . . , \mu_d)^T$ as
\begin{equation}
	\mu_i =  \max\{ 0,\min\{ 1, m_i + \epsilon \} \}, 
\end{equation}
where $\epsilon$ is sampled from  $\mathcal{N}({0,{\sigma ^2}})$. With this approximation, the objective function of the invariant mask generation module can be written as %\mu
{\small
\begin{equation}
\begin{aligned}
		\mathcal{L}_{\text {mask }}&=w_{ERM}\mathbb{E}_{e \in {\mathcal E}} \mathcal{L}^e+w_{IRM}\left\|\operatorname{Var}_{e \in {\mathcal E}}\left(\nabla_{\Theta^{\text {mask }}} \mathcal{L}^{\mathbf{e}}\right) \odot \mu \right\|^2+\frac{\lambda}{2}\|\mathbf{m}\|^2\\
		&=w_{ERM}\mathcal{L}_{ERM}+w_{IRM}\mathcal{L}_{IRM}+\frac{\lambda}{2}\|\mathbf{m}\|^2,\end{aligned}\label{eq:lmask}\end{equation}}where $\lambda$ represents the weight of the regularization term, $w_{ERM}$ and $w_{IRM}$ represent the weights of $\mathcal{L}_{ERM}$ and  $\mathcal{L}_{IRM}$, respectively. The first term is the ordinary recommended loss, which is the average loss within environment $\mathcal{E}$ and can be viewed as the ERM loss, i.e.,
%$\mathcal{L}_{ERM}$
\begin{equation}
	{\mathcal{L}_{ERM}} = \mathcal{L}( {{\Gamma^{mask}}( {u,i, {\mu  \odot \mathbf{h}_{i}}} | \Theta^{mask}  )| {\mathcal R_e^{tr}}} ), 
\end{equation}
where $\mathbf{h}_{i}$ symbolizes the weighted representations, $\Theta^{mask}$ denotes the parameters of $\Gamma^{mask}$, $\odot$ means dot product operation.
The second term is the cross-environment constraint, which is the IRM loss. The last term is the regularization term. %$\odot$ means dot product operation, 

To learns invariant masks based on the Pareto-optimal solution, architecturally, instead of just using invariant representations \cite{du2022invariant}, we incorporate an attention mechanism, which empowers us to dynamically assign weights to both the invariant representations $\mathbf{\Phi} _i$ and variant representations ${\mathbf{\Psi} _i}$. 
This attention mechanism allows our model to focus on the most relevant representations from both invariant and variant representations. Formally, the weighted representations $\mathbf{h}_i$ can be expressed as
%\vspace{-5pt}
\begin{equation}
	{\mathbf{h}_i} = \alpha _i^\Phi  \cdot {\mathbf{\Phi} _i} + \alpha _i^\Psi  \cdot {\mathbf{\Psi} _i},
\end{equation}
%\vspace{-3pt}
% to fit the weights of IRM and ERM losses in the loss function from pure collaborative filtering and content-based collaborative filtering
where $\alpha^{\Phi}_i$ and $\alpha^{\Psi}_i$ are implemented using multi-layer perceptron (MLP).  Specifically, we first concatenate the collaborative embedding and content representations of users and items, and then use two MLPs, respectively, to obtain the weights of the variant and invariant representations, which can be formalized as
\begin{equation}
	\begin{aligned}
    \alpha^{\Phi}_i=\text{MLP}_1([\mathbf{p}_u^{(t)}, \mathbf{p}_u^{(f)}, \mathbf{t}_i, \mathbf{f}_i]),\\ \alpha^{\Psi}_i=\text{MLP}_2([\mathbf{p}_u^{(t)}, \mathbf{p}_u^{(f)}, \mathbf{t}_i, \mathbf{f}_i]),	
	\end{aligned}
\end{equation}
where $\mathbf{t}_i$ and $\mathbf{f}_i$ denote the collaborative and raw multimedia representations of item $i$, and  $\mathbf{p}_u^{(t)}$ and $\mathbf{p}_u^{(f)}$ denote the corresponding user representations. In such case, the recommendation model $\Gamma( {u,i,\mathbf{h}_i} )$ can be formalized as
\begin{equation}
	\Gamma( {u,i,\mathbf{h}_i} ) = \Gamma( {\mathbf{p}_u^{(t)},\mathbf{p}_u^{(f)},{\mathbf{t}_i}, \mathbf{h}_i} ) = \langle {\mathbf{p}_u^{(t)},{\mathbf{t}_i}}\rangle  + \langle {\mathbf{p}_u^{(f)}, \mathbf{W} \cdot \mathbf{h}_i} \rangle,  
\end{equation}
where $\mathbf{W}$ refers to a projection matrix that is used to compress the dimension of the raw multimedia representations. To obtain the Pareto optimal invariant mask, we require to solve the minimization problem of the loss function $\mathcal{L}_{mask}$ via an adaptive manner, where
% and $\mathbf{W}$ is a projection matrix to compress the dimension of the raw multimedia representation.
\begin{equation}
	\begin{aligned}
		\min_{w_{ERM}, w_{IRM}}&\left\| w_{ERM} 	\nabla_{\mathbf{m}}\mathcal{L}_{ERM}+ w_{IRM} \nabla_{\mathbf{m}} \mathcal{L}_{IRM} \right\|_2^2, \\
	\text{s.t.}  \quad \text{   }& ~~  w_{ERM}+w_{IRM}=1, w_{ERM} \geq 0, w_{IRM} \geq 0,
	\end{aligned}\label{16}	
\end{equation}
with an analytical solution 
\begin{equation}
	{w}^*_{ERM}=\frac{(\nabla_{\mathbf{m}} {\mathcal{L}}_{IRM}(\mathbf{m})-\nabla_{\mathbf{m}} {\mathcal{L}}_{ERM}(\mathbf{m}))^{\top} \nabla_{\mathbf{m}} {\mathcal{L}}_{IRM}(\mathbf{m})}{\left\|\nabla_{\mathbf{m}} {\mathcal{L}}_{ERM}(\mathbf{m})-\nabla_{\mathbf{m}} {\mathcal{L}}_{IRM}(\mathbf{m})\right\|_2^2},
\end{equation}
and we clip ${w}^*_{ERM}$ to ensure  $0 \le 
 {w}^*_{ERM} \le 1$ after each iteration
 \begin{equation}
     {w^*_{ERM}} \leftarrow \max \{0, \min \{1, {w^*_{ERM}}\}\},
 \end{equation}
and let $ {w}_{IRM}=1-{w}^*_{ERM}$.
Finally, we update $\mathbf{m}$ by
\begin{equation}
	\mathbf{m}\leftarrow \mathbf{m} - s (w^*_{ERM} \nabla_{\mathbf{m}}\mathcal{L}_{ERM}+w^*_{IRM} \nabla_{\mathbf{m}} \mathcal{L}_{IRM}+ \lambda \mathbf{m}),
	\label{eq:updatem}
\end{equation}
where $s$ is the step-size for invariant mask update. %When $\mathcal{L}_{mask}$ is minimized, the optimal mask $\mathbf{m}$ is obtained.

To prove that the gradient-based update in Eq. (\ref{16}) and Eq. (\ref{eq:updatem}) lead to Pareto optimality, i.e., there exists no $\mathbf{m}^\prime$ such that $\mathcal{L}_{ERM}(\mathbf{m}^\prime)\leq \mathcal{L}_{ERM}(\mathbf{m})$ and $\mathcal{L}_{IRM}(\mathbf{m}^\prime)\leq \mathcal{L}_{IRM}(\mathbf{m})$, we follow \cite{desideri2012multiple,sener2018multi,lin2019Pareto} to consider the following optimization problem
\vspace{-3pt}
\begin{equation}
	\begin{aligned}
	(\Delta \mathbf{m}, \zeta)={}&{}\arg	\min_{\Delta \mathbf{m}, \zeta} 
 \zeta+\frac{1}{2}\left\| \Delta \mathbf{m} \right\|_2^2, \\
	& \text{s.t.}  \quad (\nabla_{\mathbf{m}}\mathcal{L}_{ERM})^T \Delta \mathbf{m}\leq \zeta,  (\nabla_{\mathbf{m}}\mathcal{L}_{IRM})^T \Delta \mathbf{m}\leq \zeta.
	\end{aligned}\label{opt}	
\end{equation}
Then we claim the solution to this optimization problem is either $\Delta \mathbf{m}=0$ and the resulting
point satisfies the Karush-Kuhn-Tucker (KKT) conditions (i.e., no other solution in its neighborhood can have lower values in both $\mathcal{L}_{ERM}$ and $\mathcal{L}_{IRM}$, thus if we want to improve the performance for a specific task, the other task's performance will be deteriorated), or the solution gives a descent direction that improves both IID and OOD generalization by reducing $\mathcal{L}_{ERM}$ and $\mathcal{L}_{IRM}$ simultaneously. 

In fact, the Lagrange function of Eq. (\ref{opt}) can be written as
\vspace{-4pt}
\begin{equation}\label{lag}
\begin{aligned}
&\mathcal{L}(\Delta \mathbf{m}, \zeta, w_{ERM}, w_{IRM})=\zeta+\frac{1}{2}\left\| \Delta \mathbf{m} \right\|_2^2\\
&+w_{ERM}((\nabla_{\mathbf{m}}\mathcal{L}_{ERM})^T \Delta \mathbf{m}- \zeta)+w_{IRM}((\nabla_{\mathbf{m}}\mathcal{L}_{IRM})^T \Delta \mathbf{m}- \zeta),
\end{aligned}
\end{equation}
where $w_{ERM}\geq 0$ and $w_{IRM}\geq 0$ are the Lagrange multipliers. Then
\begin{equation}
\begin{aligned}
\frac{\partial \mathcal{L}}{\partial \Delta_{\mathbf{m}}}&=~\Delta_{\mathbf{m}}-w_{ERM} \cdot \nabla_{\mathbf{m}}\mathcal{L}_{ERM}-w_{IRM}\cdot \nabla_{\mathbf{m}} \mathcal{L}_{IRM}=0,\\
& \Rightarrow \Delta_{\mathbf{m}}=-w_{ERM} \cdot \nabla_{\mathbf{m}}\mathcal{L}_{ERM}-w_{IRM}\cdot \nabla_{\mathbf{m}} \mathcal{L}_{IRM}, \\
\frac{\partial \mathcal{L}}{\partial \zeta}&=~1-w_{ERM}-w_{IRM},\quad \Rightarrow w_{ERM}+w_{IRM}=1.
\end{aligned}
\end{equation}
Notably, the dual problem of Eq. (\ref{opt}) is Eq. (\ref{16}), and according to KKT condition, we have
\begin{equation}
\begin{aligned}
w^*_{ERM}((\nabla_{\mathbf{m}}\mathcal{L}_{ERM})^T \Delta \mathbf{m}^*- \zeta^*)&=0,\\
w^*_{IRM}((\nabla_{\mathbf{m}}\mathcal{L}_{IRM})^T \Delta \mathbf{m}^*- \zeta^*)&=0.
\end{aligned}
\end{equation}
Thus, if $\Delta \mathbf{m}^*=0$, then  $(\nabla_{\mathbf{m}}\mathcal{L}_{ERM})^T \Delta \mathbf{m}^*=(\nabla_{\mathbf{m}}\mathcal{L}_{IRM})^T \Delta \mathbf{m}^*=0$. If $\Delta \mathbf{m}^*\neq0$, then we have $-\left\| \Delta \mathbf{m}^* \right\|_2^2-\zeta^*=0$, which implies that $(\nabla_{\mathbf{m}}\mathcal{L}_{ERM})^T \Delta \mathbf{m}^*\leq \zeta^*\leq -\left\| \Delta \mathbf{m}^* \right\|_2^2$ and $(\nabla_{\mathbf{m}}\mathcal{L}_{IRM})^T \Delta \mathbf{m}^*\leq \zeta^*\leq -\left\| \Delta \mathbf{m}^* \right\|_2^2$, and reduces $\mathcal{L}_{ERM}$ and $\mathcal{L}_{IRM}$ simultaneously.
%\vspace{-4pt}
\subsection{Representation Convertion}
\label{section: 3.5}
Based on the invariant mask obtained by the invariant mask generation module, we use the convert module to divide the raw multimedia representations into variant representations and invariant representations. Specifically, the invariant representations are
\begin{equation}
	\mathbf{\Phi}_i=\mathbf{m} \odot \mathbf{f}_i. 
\end{equation}
Correspondingly, the variant representations can be expressed as
\begin{equation}
	\mathbf{\Psi}_i=(1-\mathbf{m}) \odot \mathbf{f}_i, 
\end{equation}
where $\mathbf{m} \in [0,1]^d$ is the float invariant mask. %The final set of invariant and variant representations are denoted %The final set of invariant and variant representations are denoted as $\left\{\mathbf{\Phi}_i \mid i \in \mathcal{I}\right\}$ and $\left\{\mathbf{\Psi}_i \mid i \in \mathcal{I}\right\}$, respectively. 
%Obviously, the core of this module is the learning of invariant masks $\mathbf{m}$.

%The whole framework is jointly optimized, so that the mutual promotion between HIM, invariant mask generation module, and convert module can be fully leveraged.
%\vspace{-4pt}
\subsection{Final Recommendation Model}
By repeating $T$ times the workflow shown in Figure \ref{fig:framework} until convergence, stable invariant masks are generated. Thus, we learn the final recommendation model $\Gamma^*(u, i, \mathbf{\Phi} _i | \Theta^*)$ parameterized by $\Theta^*$ based on the invariant representations generated by the convert module. The learning objective  shown in Eq. (\ref{eq:problemF}) can be rewritten as
\begin{equation}
	\arg \min _{{\Theta ^*}} \mathcal{L}(\Gamma^*(u, i, \mathbf{\Phi} _i | \Theta^*)| {\mathcal R^{tr}}).
	\label{eq:final}
\end{equation}
The whole training process of PaInvRL is described in Algorithm \ref{alg1}.
%\vspace{-3pt}
\begin{algorithm}[t]
	\caption{The overall training process of PaInvRL.}
	\label{alg1}
	\LinesNumbered 
	\KwIn{$\mathcal{R^{+}}, \mathcal{R}^{-}, \mathcal{R}^{t r}$.}
	%\KwOut{output result}
	%some description\; 
	\For{$i \leftarrow 1$ to $T$}{
		\While{not converge}{
			\For{$e \in \mathcal{E}$}{
				Optimize $\Gamma_{(e)}$ via Eq. (\ref{eq:HIM})\;
			}
			\For{$e \in \mathcal{E}$}{
				Compute $\mathcal{R}_e$ via Eq. (\ref{eq:IP})\;
			}
	}
	\While{not converge}{
		% Obtain $\hat{w}^*_{ERM}$, $\hat{w}_{IRM}$ using the analytical solution form\;
  ${w}^*_{ERM}=\frac{(\nabla_{\mathbf{m}} {\mathcal{L}}_{IRM}(\mathbf{m})-\nabla_{\mathbf{m}} {\mathcal{L}}_{ERM}(\mathbf{m}))^{\top} \nabla_{\mathbf{m}} {\mathcal{L}}_{IRM}(\mathbf{m})}{\left\|\nabla_{\mathbf{m}} {\mathcal{L}}_{ERM}(\mathbf{m})-\nabla_{\mathbf{m}} {\mathcal{L}}_{IRM}(\mathbf{m})\right\|_2^2}$\;
		${w^*_{ERM}} \leftarrow \max \{0, \min \{1, {w^*_{ERM}}\}\}$\;
  $ {w}^*_{IRM}=1-{w}^*_{ERM}$\;
  $\mathbf{m}\leftarrow \mathbf{m} - s (w^*_{ERM} \nabla_{\mathbf{m}}\mathcal{L}_{ERM}+w^*_{IRM} \nabla_{\mathbf{m}} \mathcal{L}_{IRM}+ \lambda \mathbf{m})$\;
%		$\mathbf{m}\leftarrow \mathbf{m} - \eta (\hat w_\mathrm{ERM} \nabla_{\mathbf{m}}\mathcal{L}_\mathrm{ERM}+\hat  w_\mathrm{IRM} \nabla_{\mathbf{m}} \mathcal{L}_\mathrm{IRM} + \lambda \mathbf{m})$\;
	}}
	Optimize $\Gamma^*(u, i, \Phi | \Theta^*)$ via Eq. (\ref{eq:final})\;
	\KwOut{Final recommendation model $\Gamma^*(u, i, \mathbf{\Phi} |\Theta^*)$.}

\end{algorithm}

\newpage
\section{Experiments}
\label{section: 4}
In this section, we conduct experiments on three widely used real-world datasets to answer the following research questions:
\begin{itemize}
	\item {\textbf{RQ1}}: Can PaInvRL outperform other recommendation methods in both IID and OOD tasks?
	\item {\textbf{RQ2}}: How masks incorporating attention mechanisms affect learned representations? 
	\item {\textbf{RQ3}}: How does each component in $\mathcal{L}_{\text {mask}}$ affect the performance of PaInvRL in both IID and OOD tasks? 
        \item {\textbf{RQ4}}: How does the number of environments affect the performance of PaInvRL?
\end{itemize}

\subsection{Datasets}
% \subsubsection{Datasets.}
We conduct experiments on three publicly available real-world datasets: Movielens, Tiktok, and Kwai. The summary statistics of these datasets are shown in Table \ref{tab:dataset}.\\
\begin{table}[tbp]
	\centering
 \setlength{\tabcolsep}{2.5pt}
	\caption{The statistics of datasets. $d_V$, $d_A$, and $d_T$ denote the dimensions of visual, acoustic, and textual modalities.}
        \vspace{-9pt}
	\resizebox{1.0\linewidth}{!}{
	\begin{tabular}{lrrrrrrr}
		\toprule
		\textbf{Dataset} & {\textbf{\#Interactions}} & {\textbf{ \#Items}} & {\textbf{ \#Users}} & {\textbf{ Sparsity}} & \textbf{$d_V$} & \textbf{$d_A$} & \textbf{$d_T$} \\
		\midrule
		\textbf{Movielens } & 1,239,508 & 5,986 & 55,485 & 99.63\% & 2,048 & {128} & \multicolumn{1}{r}{100} \\
		\textbf{Tiktok } & 726,065 & 76,085 & 36,656 & 99.99\% & 128   & {128} & {128} \\
		\textbf{Kwai } & 298,492 & 86,483 & 7,010 & 99.98\% & 2,048 &  -    & - \\
		\bottomrule
	\end{tabular}%
}
	\label{tab:dataset}%
\end{table}%
\textbf{Movielens.} This dataset is widely used in personalized recommendation tasks. The dataset is constructed by collecting movie titles and descriptions from the Movielens dataset\footnote{https://movielens.org/.} and retrieving the corresponding trailers. The visual, acoustic, and textual representations were extracted from the pre-trained ResNet50 \cite{he2016deep}, VGGish \cite{hershey2017cnn}, and Sentence2Vec \cite{arora2017simple}, respectively. \\
\textbf{Tiktok.} It is collected from the micro-video sharing platform TikTok\footnote{https://www.tiktok.com/.}. It includes micro-videos with a duration of 3-15 seconds, along with video captions, user information, and user-item interactions. The multi-modal representations include visual, acoustic, and textual representations of micro-videos. All of the multi-modal representations are provided by the official.\\
\textbf{Kwai. } It is a large-scale micro-video dataset collected from the Kwai platform\footnote{https://www.kwai.com/.}. Similar
to the TikTok dataset, it includes user information, micro-video content representations, and interaction data. We follow the previous work \cite{du2022invariant} to obtain the raw multimedia representations. It should be noticed that this dataset only includes visual representations.
%\begin{table}
%	\caption{Frequency of Special Characters}
%	\label{tab:freq}
%	\begin{tabular}{ccl}
%		\toprule
%		Non-English or Math&Frequency&Comments\\
%		\midrule
%		\O & 1 in 1,000& For Swedish names\\
%		$\pi$ & 1 in 5& Common in math\\
%		\$ & 4 in 5 & Used in business\\
%		$\Psi^2_1$ & 1 in 40,000& Unexplained usage\\
%		\bottomrule
%	\end{tabular}
%\end{table}
\begin{table*}[htbp]
	\centering
	\caption{Performance comparison  on different datasets in terms of Recall@10, Precision@10, and NDCG@10.}
        \vspace{-9pt}  
	\begin{tabular}{clcccccccccc}
		\toprule
		\multirow{2}[4]{*}{\textbf{Task}} & \multicolumn{1}{c}{\multirow{2}[4]{*}{\textbf{Method}}} & \multicolumn{1}{c}{\multirow{2}[4]{*}{\textbf{Modality}}} & \multicolumn{3}{c}{\textbf{Movielens}} & \multicolumn{3}{c}{\textbf{Tiktok}} & \multicolumn{3}{c}{\textbf{Kwai}} \\
		\cmidrule{4-12}          &       &   & \multicolumn{1}{l}{\textbf{P@10}} & \multicolumn{1}{l}{\textbf{R@10}} & \multicolumn{1}{l}{\textbf{N@10}} & \multicolumn{1}{l}{\textbf{P@10}} & \multicolumn{1}{l}{\textbf{R@10}} & \multicolumn{1}{l}{\textbf{N@10}} & \multicolumn{1}{l}{\textbf{\textbf{P@10}}} & \multicolumn{1}{l}{\textbf{R@10}} & \multicolumn{1}{l}{\textbf{N@10}} \\
		\midrule
        \multirow{8}[2]{*}{\textbf{IID}} & 	\textbf{NGCF \cite{wang2019neural}}  & Single & 0.0180  & 0.1355  & 0.0383  & 0.0138  & 0.0409  & 0.0513  & 0.0425  & 0.0487  & 0.0697  \\
		& \textbf{UltraGCN \cite{mao2021ultragcn}} & Single & 0.0126  & 0.1060  & 0.0418  & 0.0163  & 0.0437  & 0.0543  & 0.0459  & 0.0509  & 0.0729  \\
		& \textbf{LightGCN \cite{he2020lightgcn}} & Single & 0.0215  & 0.1643  & 0.0554  & 0.0164  & 0.0444  & \textbf{0.0584}  & 0.0496  & 0.0435  & 0.0688  \\
        &  \textbf{VBPR \cite{he2016vbpr}}  & Multi & 0.0176  & 0.1290  & 0.0400  & 0.0142  & 0.0409  & 0.0469  & 0.0409  & 0.0476  & 0.0682  \\
		& \textbf{MMGCN \cite{wei2019mmgcn}} & Multi & 0.0207  & 0.1613  & 0.0641  & 0.0154  & 0.0444  & \textbf{0.0584}  & 0.0496  & 0.0535  & 0.0738  \\
		& \textbf{InvRL \cite{du2022invariant}} & Multi & 0.0218  & \textbf{0.1681} & 0.0617  & 0.0213  & 0.0440  & 0.0576  & 0.0528  & 0.0549  & 0.0729  \\
        & \textbf{MMSSL \cite{wei2023multi}} & Multi & 0.0237  & 0.1587  & 0.0572  & 0.0202  & 0.0443  & 0.0555  & 0.0523  & 0.0518  & 0.0748  \\
		& \textbf{PaInvRL (ours)} & Multi & \textbf{0.0240} & 0.1660  & \textbf{0.0650} & \textbf{0.0229} & \textbf{0.0463} & 0.0578 & \textbf{0.0536} & \textbf{0.0595} & \textbf{0.0815} \\
        \midrule
		\multicolumn{1}{r}{\multirow{8}[2]{*}{\textbf{OOD}}} &
         \textbf{NGCF \cite{wang2019neural}} & Single & 0.0191 & 0.0733 & 0.0474 & 0.0048 & 0.0060 & 0.0153 & 0.0411 & 0.0828 & 0.1466 \\
		& \textbf{UltraGCN \cite{mao2021ultragcn}} & Single & 0.0212 & {0.0708} & 0.0508 & 0.0043 & 0.0061 & 0.0174 & 0.0321 & 0.0784 & 0.1464 \\
		& \textbf{LightGCN \cite{he2020lightgcn}} & Single & 0.0159 & 0.0638 & 0.0412 & 0.0053 & 0.0082 & 0.0169 & 0.0420 & 0.0883 & 0.1331 \\
        & \textbf{VBPR \cite{he2016vbpr}}  & Multi & 0.0165 & 0.0649 & 0.0396 & 0.0036 & 0.0057 & 0.0153 & 0.0324 & 0.0763 & 0.1504 \\
		& \textbf{MMGCN \cite{wei2019mmgcn}} & Multi & 0.0188 & 0.0732 & 0.0463 & 0.0044 & 0.0058 & 0.0161 & 0.0402 & 0.0831 & 0.1503 \\
		& \textbf{InvRL \cite{du2022invariant}} & Multi & 0.0253 & 0.0791 & 0.0543 & 0.0059 & 0.0097 & 0.0226 & 0.0407 & 0.0862 & 0.1894 \\
        & \textbf{MMSSL \cite{wei2023multi}} & Multi & 0.0225 & 0.0813 & 0.0537 & 0.0055 & 0.0098 & 0.0234 & 0.0462 & 0.1036 & 0.1682 \\
		& \textbf{PaInvRL (ours)} & Multi & \textbf{0.0291} & \textbf{0.0825} & \textbf{0.0584} & \textbf{0.0067} & \textbf{0.0107} & \textbf{0.0252} & \textbf{0.0524} & \textbf{0.1113} & \textbf{0.2061} \\
		\bottomrule
	\end{tabular}%
	\label{tab:preformence}%
\end{table*}%
\begin{figure*}[t]
	\centering
	\subfigure[Movielens]{
		\includegraphics[width=0.32\linewidth]{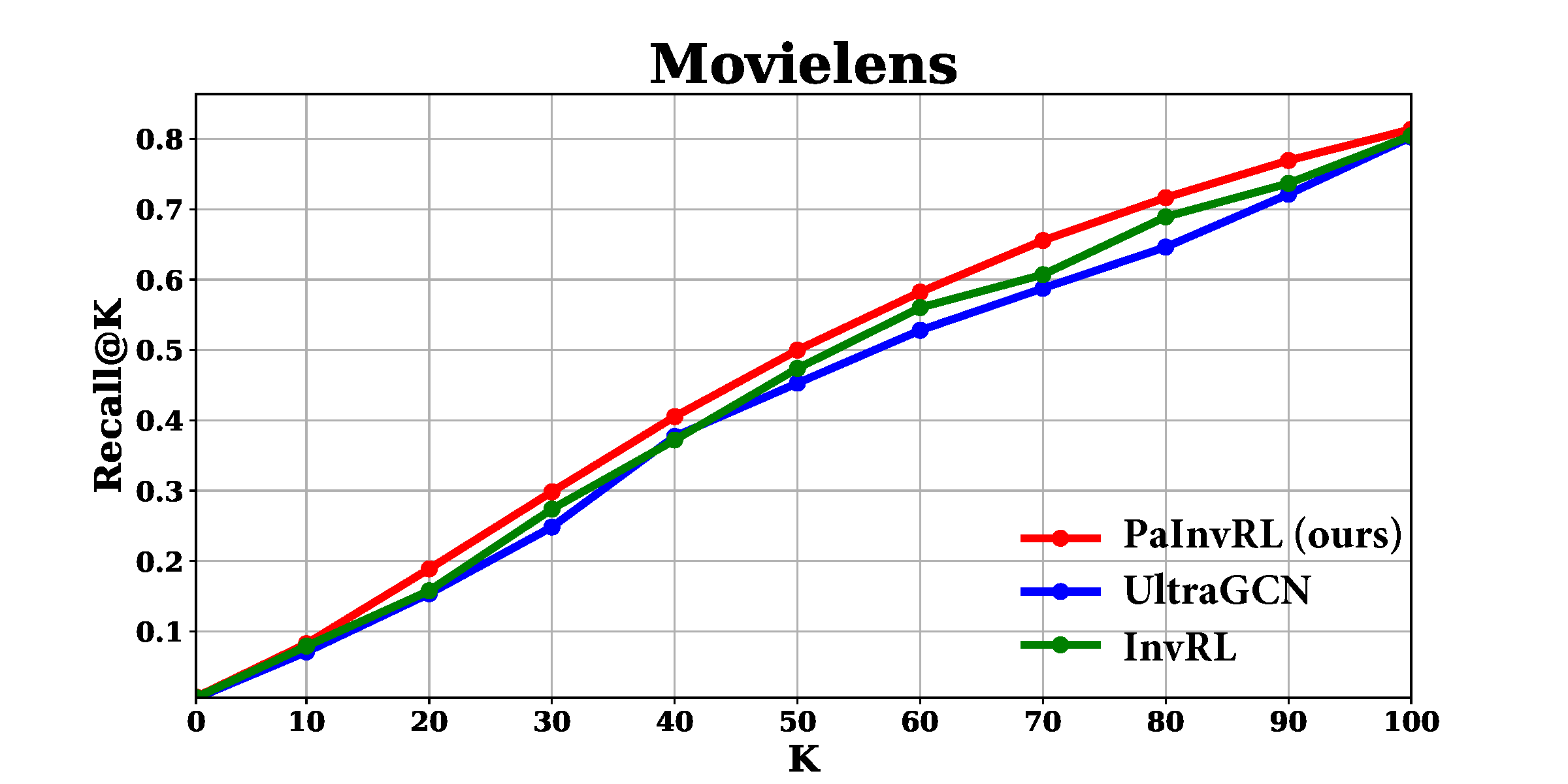}
	}
	\subfigure[Tiktok]{
		\includegraphics[width=0.32\linewidth]{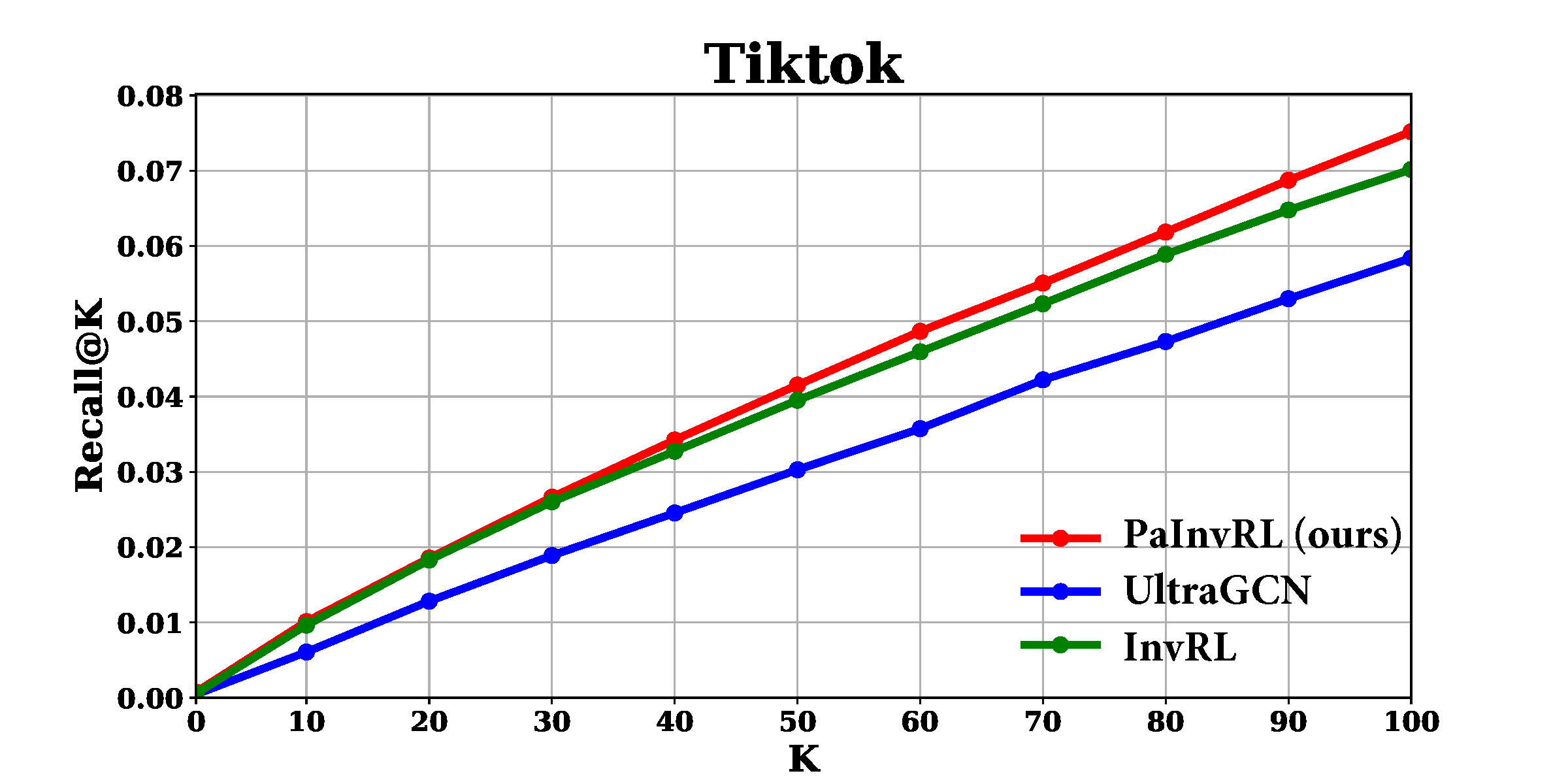}
	}
	\subfigure[Kwai]{
		\includegraphics[width=0.32\linewidth]{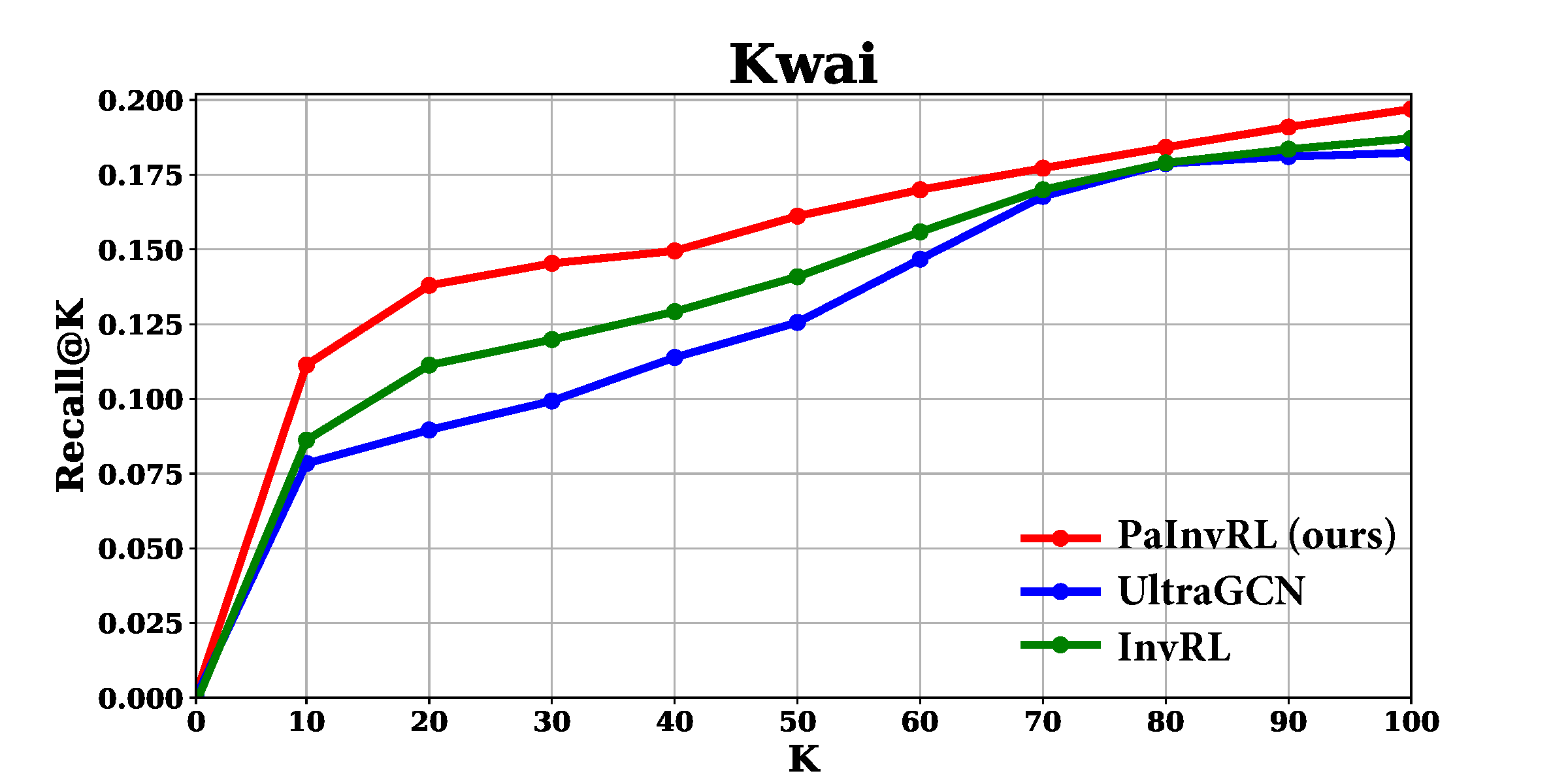}
	}
        \vspace{-9pt}
	\caption{The performance comparison between UltraGCN, InvRL and PaInvRL on all three datasets using Recall$@$K as evaluation metric, where K is varying from 0 to 100 with step-size 10.}
 \vspace{-6pt}
	\label{fig:Recall}
\end{figure*}

\subsection{Experiment Setup}

\subsubsection{Baselines}
To verify the effectiveness of PaInvRL, we compare it with the following baseline methods: \\
\textbf{NGCF \cite{wang2019neural}.} It is based on graph neural networks that explicitly encode collaborative signals as higher-order connections by performing embedding propagation. \\
\textbf{UltraGCN \cite{mao2021ultragcn}.} It is an ultra-simplified GCN model that does not perform explicit message passing, but directly approximates the limit of infinite layer graph convolutions by constraining losses.  \\
\textbf{LightGCN \cite{he2020lightgcn}.} It is a graph-based model designed to improve the performance and efficiency of recommendations by simplifying the graph convolution networks. \\
\textbf{VBPR \cite{he2016vbpr}.} It is the first model that considers introducing visual representation into the recommendation system by concatenating visual embeddings with id embeddings as the item representations.\\
\textbf{MMGCN \cite{wei2019mmgcn}.} It is a model that builds on the message-passing idea of graph neural networks to generate user and micro-video-specific pattern representations to capture user preferences better.  \\
%Dual-GNN \cite{wang2021dualgnn}. This framework is based on the user-microvideo bilateral graph and the user co-occurrence graph, and exploits the correlation between users to collaboratively mine each user's specific fusion patterns.  \\
\textbf{InvRL \cite{du2022invariant}.} This model introduces IRM to multi-modal recommendations for the first time, which mitigates the effects of spurious correlations by learning invariant item representations.   \\
\textbf{MMSSL \cite{wei2023multi}.} This method solves the problem of label sparsity in multimedia recommendations by two-stage self-supervised learning to achieve modality-aware data scaling.%GRCN \cite{wei2020graph},
\begin{figure*}[t]
	\centering
	\subfigure[Movielens]{
		\includegraphics[width=0.32\linewidth]{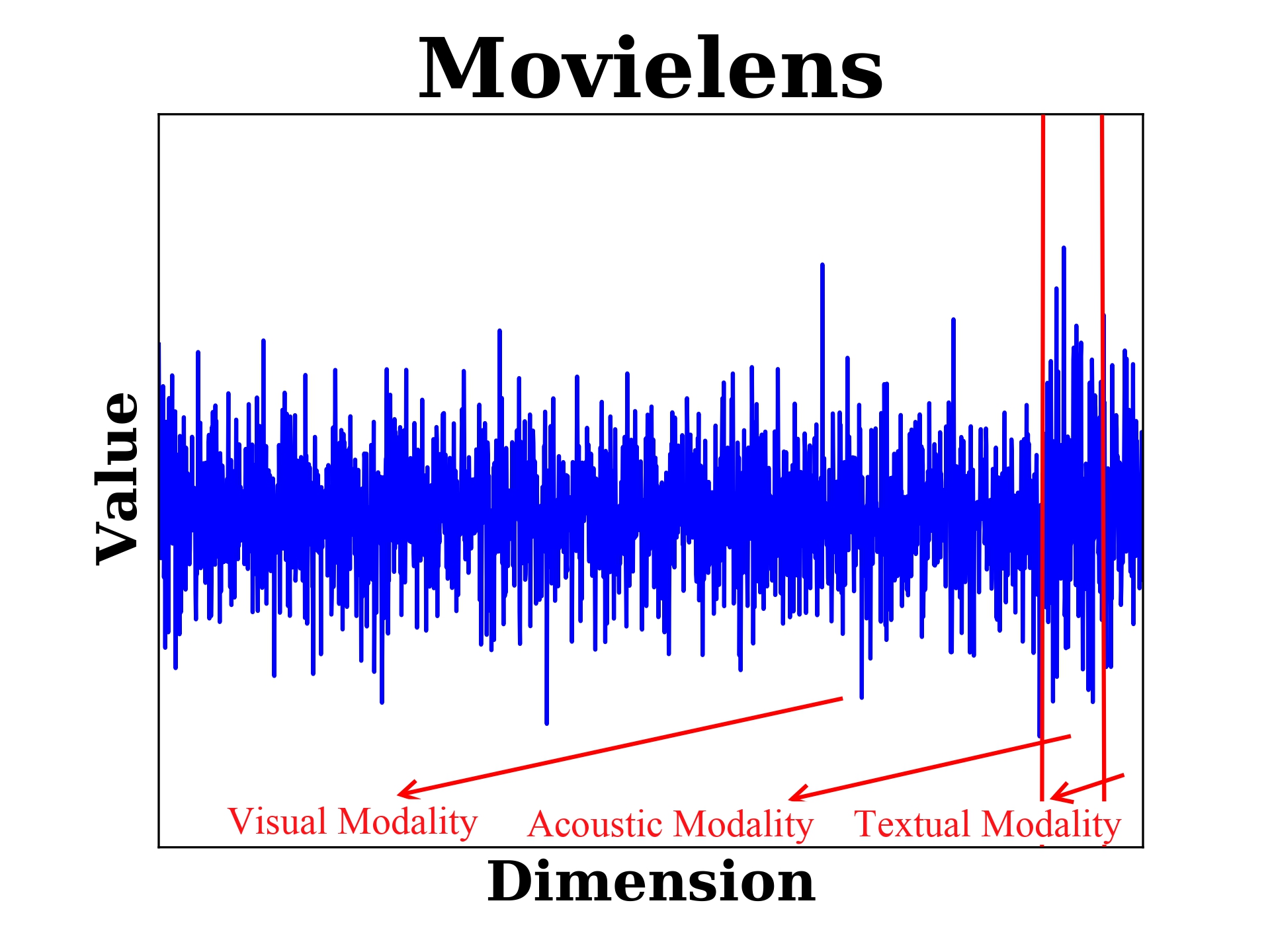}
	}
	\subfigure[Tiktok]{
		\includegraphics[width=0.32\linewidth]{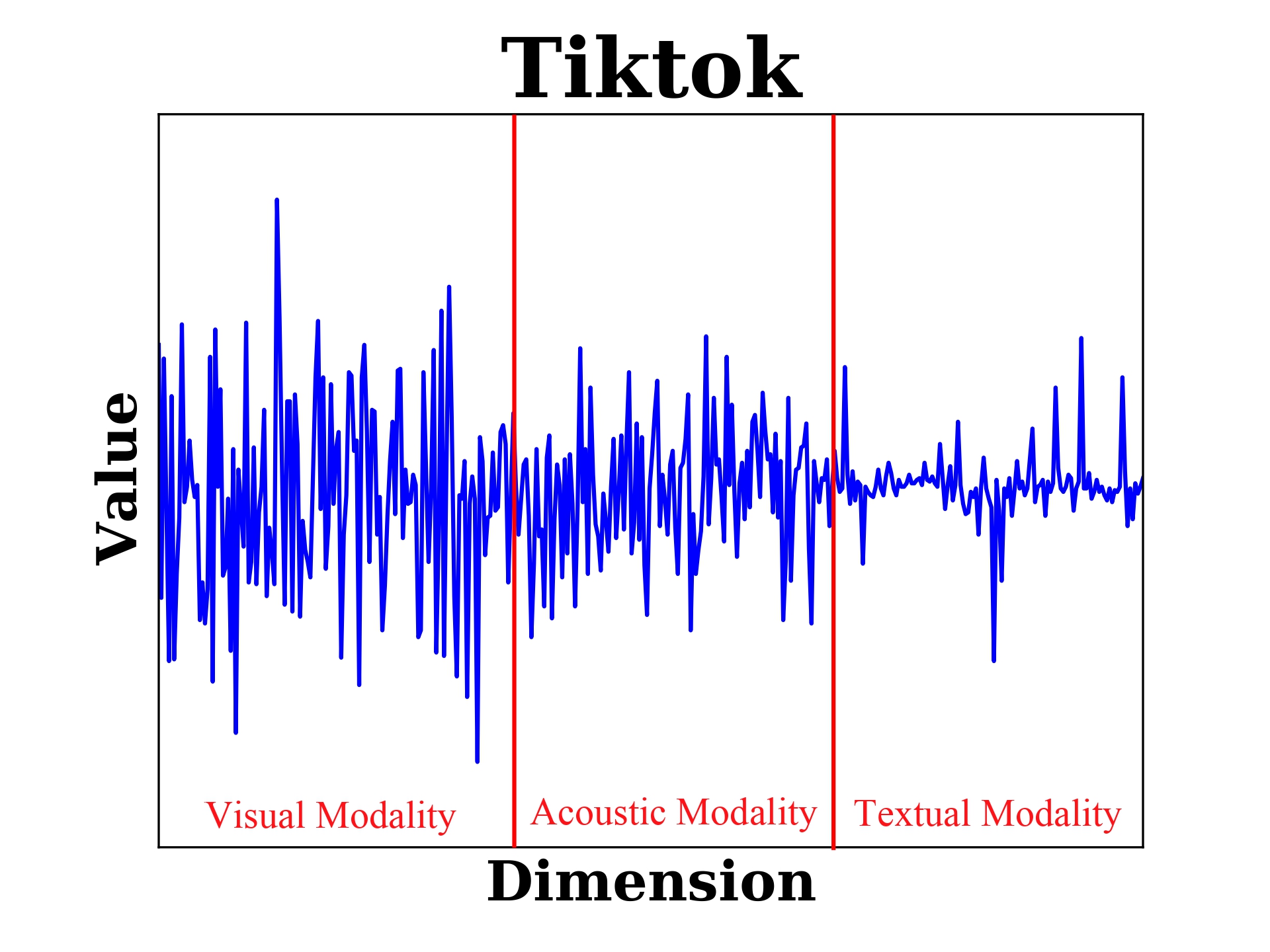}
	}
	\subfigure[Kwai]{
		\includegraphics[width=0.32\linewidth]{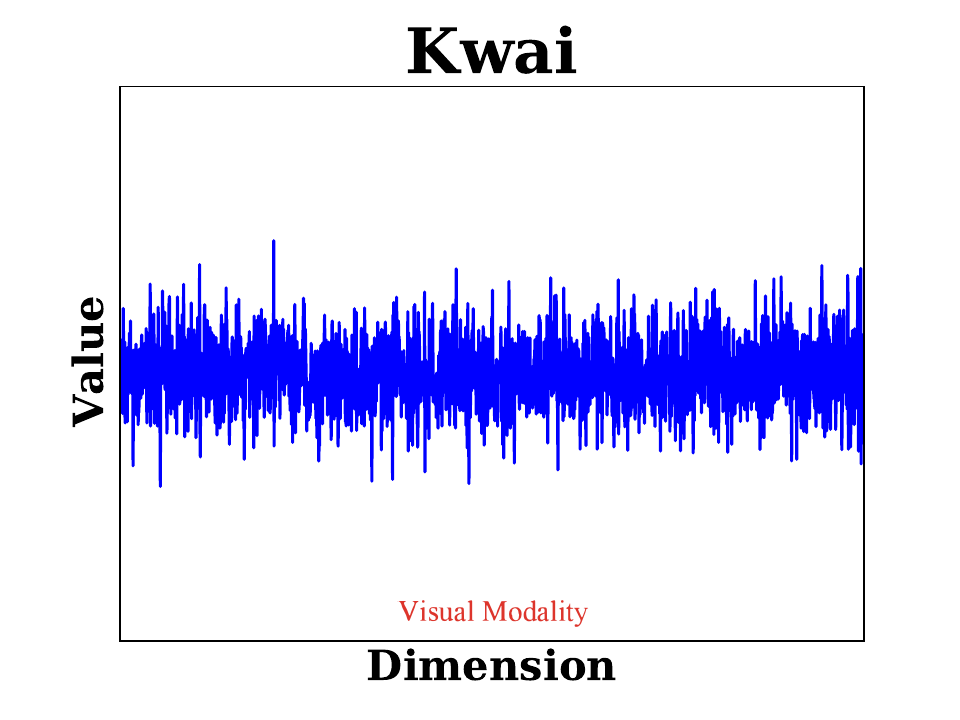}
	}
  \vspace{-9pt}
	\caption{Visualization of the masks on different modalities and corresponding patterns on all three datasets.}
	\label{fig:mask}
\end{figure*}
\begin{table*}[htbp]
	\centering
	\caption{Performance comparison with different loss components in the IID and OOD tasks.}
         \vspace{-9pt}
	\begin{tabular}{clrrrrrrrrr}
		\toprule
		\multirow{2}[4]{*}{\textbf{Task}} & \multicolumn{1}{c}{\multirow{2}[4]{*}{\textbf{Loss}}} & \multicolumn{3}{c}{\textbf{Movielens}} & \multicolumn{3}{c}{\textbf{Tiktok}} & \multicolumn{3}{c}{\textbf{Kwai}} \\
		\cmidrule{3-11}          &       & \multicolumn{1}{c}{\textbf{P@10}} & \multicolumn{1}{c}{\textbf{R@10}} & \multicolumn{1}{c}{\textbf{N@10}} & \multicolumn{1}{c}{\textbf{P@10}} & \multicolumn{1}{c}{\textbf{R@10}} & \multicolumn{1}{c}{\textbf{N@10}} & \multicolumn{1}{c}{\textbf{P@10}} & \multicolumn{1}{c}{\textbf{R@10}} & \multicolumn{1}{c}{\textbf{N@10}} \\
		\midrule
		\multirow{3}[1]{*}{\textbf{IID}} & \textbf{$\mathcal L_{ERM}$} & 0.0253  & 0.1068  & 0.0410  & 0.0263  & 0.0537  & 0.0643  & 0.0574  & 0.0619  & 0.0837  \\
		& \textbf{$\mathcal L_{IRM}$} & 0.0141  & 0.1027  & 0.0311  & 0.0217  & 0.0385  & 0.0572  & 0.0562  & 0.0612  & 0.0831  \\
		& \textbf{$\mathcal L_{ERM}+ \mathcal L_{IRM}$} & 0.0240  & 0.1660  & 0.0650  & 0.0229  & 0.0463  & 0.0578  & 0.0536  & 0.0595  & 0.0815  \\
  \midrule
		\multirow{3}[1]{*}{\textbf{ OOD}} & \textbf{$\mathcal L_{ERM}$} & 0.0212  & 0.0583  & 0.0380  & 0.0058  & 0.0097  & 0.0234  & 0.0508  & 0.1026  & 0.1983  \\
		& \textbf{$\mathcal L_{IRM}$} & 0.0353  & 0.1009  & 0.0693  & 0.0068  & 0.0103  & 0.0235  & 0.0575  & 0.1504  & 0.2522  \\
		& \textbf{$\mathcal L_{ERM}+ \mathcal L_{IRM}$} & 0.0291  & 0.0825  & 0.0584  & 0.0067  & 0.0107  & 0.0252  & 0.0524  & 0.1113  & 0.2061  \\
		\bottomrule
	\end{tabular}%
	\label{tab:lossfunction}%
\end{table*}%
\subsubsection{Experiment Protocol and Details} Following the previous work \cite{du2022invariant}, three widely-used metrics are adopted to evaluate the ranking performance: Recall$@$K (R$@$K), NDCG$@$K (N$@$K), and Precision$@$K (P$@$K). We set $K = 10$ in our experiments. All the experiments are implemented with PyTorch \cite{paszke2019pytorch} and Adam is implemented as the optimizer. The embedding size is fixed to 64 for all models. For Movielens, we set $d_V$ = 2,048, $d_A$ = 128, and $d_T$ = 10. For Tiktok, we set $d_V$ = 128, $d_A$ = 128, and $d_T$ = 128. For Kwai, we only use visual representations and set $d_V$ = 4,096. The batch size is set to 512 and the number of environments is set to 10. We also set the parameters $\lambda$ in Eq. (\ref{eq:lmask}) to 1, and the hyper-parameters $\eta $ and $\kappa$ in Eq. (\ref{eq:ultragcnloss}) to 0.0001 and 0.01, respectively. 
The heterogeneous identification module, invariant mask generation module, and the final recommendation model are trained for 20, 40, and 500 epochs, respectively.
To evaluate the performance of PaInvRL in both IID and OOD tasks, we first use UltraGCN to identify two environments using the heterogeneous identification module. We train the model in the environment that contains more data, and test the model in the environment that contains less data for the OOD task. We split the training set for the OOD task into two parts with 9:1 ratio to obtain the training set and test set for the IID task.

\subsection{Performance Comparison (RQ1)}
We report the performance of various methods on all three datasets in Table \ref{tab:preformence}, where the best-performing baselines are bolded. We have the following observations.

First, multi-modality-based methods outperform single-modality-based methods in both IID and OOD tasks, and MMSSL achieves the most competitive performance among all the baseline methods.

Second, in the OOD recommendation task, it shows that PaInvRL significantly outperforms other methods, due to PaInvRL learning invariant representations and identifying spuriously correlation. In addition, it should be noticed that PaInvRL outperforms InvRL, which is attributed to PaInvRL learning a better mask by considering Pareto optimization and weighting both invariant and variant representation using the attention mechanism.

Third, in the IID task, although InvRL achieved better performance than some single-modality-based methods like NGCF and LightGCN, its performance is not as good as that of other multi-modality-based methods like MMGCN. This is because InvRL only focuses on learning invariant representations, which leads to performance degradation in the IID task. However, the proposed method PaInvRL weights the ERM loss $\mathcal{L}_{ERM}$ and IRM loss $\mathcal{L}_{IRM}$ to ensure the learned representations are able to perform well in both IID and OOD tasks. Therefore, PaInvRL also achieves the best performance compared to other methods in the IID task.

Overall speaking, PaInvRL not only outperforms other baseline methods in the OOD task, but also has the best performance in the IID task.
In addition, we conduct a more detailed experiment to compare PaInvRL, InvRL, and UltraGCN using Recall$@$K as the evaluation metric in the OOD task on all three datasets. The results are presented in Figure \ref{fig:Recall}, which indicates that PaInvRL stably outperforms UltraGCN and InvRL across different K values, which further verifies the effectiveness of the proposed method.
% and indicates the proposed method improves IID performance without compromising OOD performance.
%\vspace{2pt}
\subsection{Ablation Study (RQ2)}
In ablation studies, we first investigate the effect of IRM loss $\mathcal{L}_{IRM}$ and ERM loss $\mathcal{L}_{ERM}$, which is used for training the invariant mask generation module. Then we discuss how the generated mask $m$ works. Additionally, we conducted experiments to study the impact of environmental quantity on experimental performance.

We consider three cases in the ablation study: only with the ERM loss $\mathcal L_{ERM}$, only with the IRM loss $\mathcal L_{IRM}$, and with adaptively weighted ERM loss and IRM loss $\mathcal L_{ERM} + \mathcal L_{IRM}$ across all three datasets. The experiment results are shown in Table \ref{tab:lossfunction}. From this table, we can observe that when PaInvRL only with the IRM loss, it is able to achieve the best performance in the OOD task but perform the worst in the IID task. Meanwhile, when only using the ERM loss, it will perform best in the IID task but perform worst in the OOD task. It shows that if only focusing on a single task, though we will obtain a competitive result, the model performance is harmed in another task, which shows the necessity of considering two tasks simultaneously. When using weighted IRM loss and ERM loss together, it has competitive performance in both IID tasks and OOD tasks. Overall, using only one type of loss or simply using a hyper-parameter to weight them directly (i.e., InvRL) cannot achieve good recommendation performance. When we adaptively weight ERM loss and IRM loss together and obtain the weights from a Pareto optimal solution, it obtains competing recommendation performance. This can be attributed to the fact that our solution cannot be dominated by other solutions. In other words, there does not exist any solution that performs better than our solution on both IID and OOD tasks at the same time. 
%\vspace{-8.8pt}
\subsection{In Depth Analysis (RQ3, RQ4)}
\begin{figure}[t]
	\centering
 \subfigure[IID]{
\includegraphics[width=0.47\linewidth]{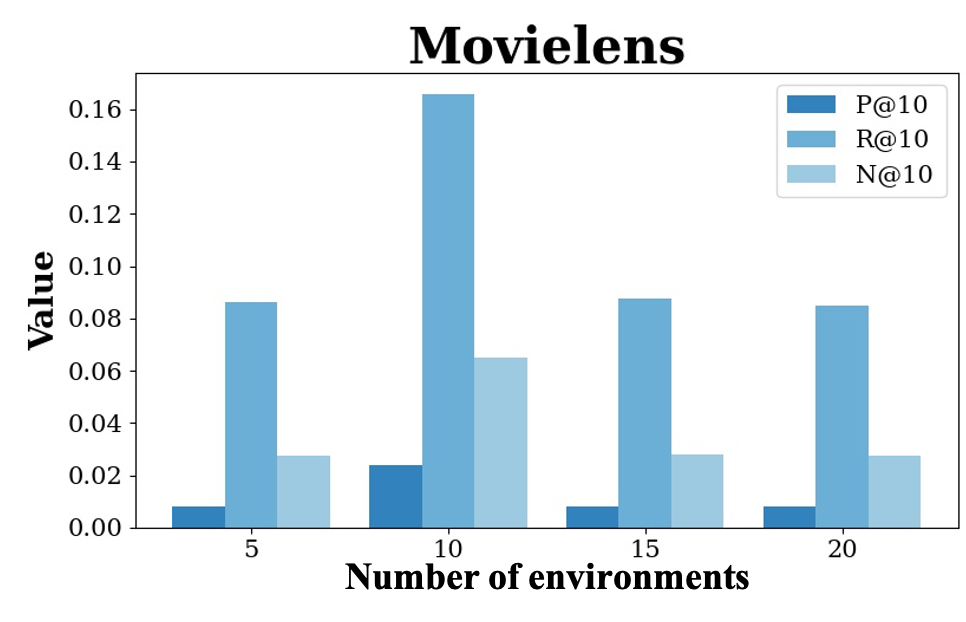}
	}
	\subfigure[OOD]{
\includegraphics[width=0.47\linewidth]{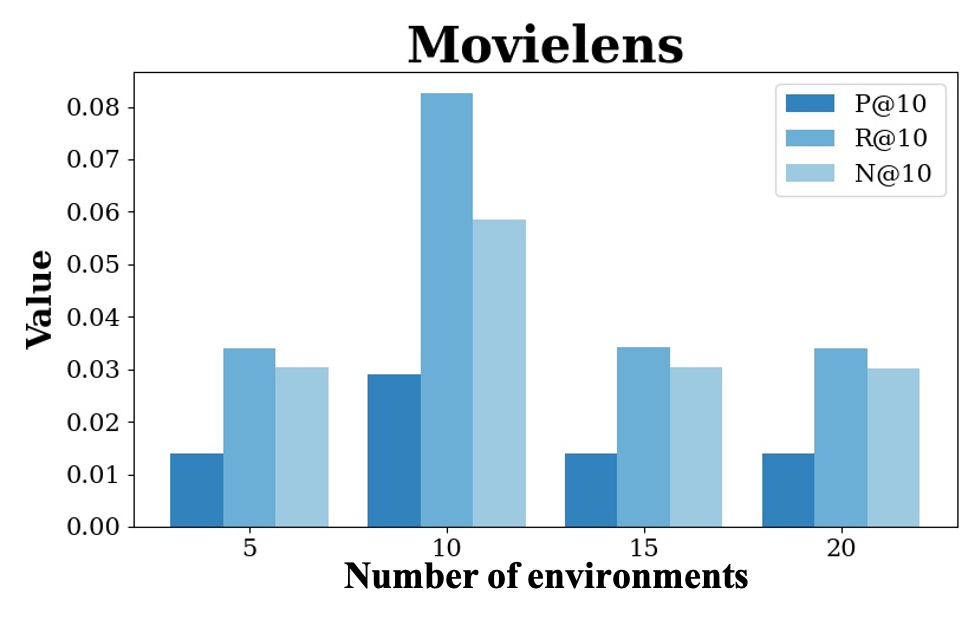}
	}
\vspace{-9pt}
	\caption{Experimental comparison of different environment numbers on IID and OOD recommendation tasks.}
	\label{fig:domain}
 \vspace{-10pt}
\end{figure}

\textbf{Study on the Generated Mask (RQ3).} 
To study the effect of the generated mask $m$ in the invariant mask generation module, we visualize the invariant mask generated on three datasets, Movielens, Tiktok, and Kwai, as shown in Figure \ref{fig:mask}. According to the results in Figure \ref{fig:mask}, the generated masks show different distributions in different modalities, especially, for the Movielens and Tiktok datasets, which contain three different modal representations of visual, acoustic, and textual. Since the Kwai dataset has only one modal representation, the distribution of the masks varies subtly. Additionally, our method demonstrates a more uniform distribution across different modalities compared to InvRL \cite{du2022invariant}. It can be attributed to the fact that PaInvRL learns a better mask by considering Pareto optimization during mask generation, while InvRL only considers a simple hyper-parameter to weight two losses together.\\
\textbf{Study on Number of Environments (RQ4).} 
To investigate the capacity of PaInvRL under different numbers of environments, we conduct several experiments on the Movielens dataset with different numbers of environments. The experimental results are shown in Figure \ref{fig:domain}. First, PaInvRL performs better under a moderate number of experiments. When the number of environments is small, we cannot effectively separate the variant and invariant information. When the number of environments is large, only a few samples are in each environment. Therefore, either too small or too large number will harm the performance of the proposed method.
\section{Conclusions}
In this paper, we provide a fresh perspective on the optimization dilemma in the IID-OOD generalization task of multimedia recommendation from a multi-objective optimization viewpoint. We propose a new Pareto-optimality-based invariant representation learning method, PaInvRL, which adaptively assigns the weights of ERM loss and IRM loss to obtain Pareto-optimal solutions. In contrast to previous approaches like InvRL, our gradient-based invariant mask generation method is shown to provide a descent direction that improves both IID and OOD generalization by reducing ERM and IRM losses simultaneously. This allows the final recommendation model trained on the learned invariant representations to achieve Pareto optimality in both IID and OOD recommendation tasks. Extensive experimental results show that our method achieves significant performance improvements compared to various baselines on three public datasets. In our future work, it is interesting to enhance the explainability of the learned invariant representations by developing a GNN-based explainer to learn causal effects on modality-aware user-item interaction graphs. This will help provide insights into how the invariant representations contribute to the recommendation performance and enable us to make more informed decisions in the recommendation process. 
\newpage
\begin{acks}
This work was supported by grants from the National Major Science and Technology Projects of China (grant no: 2022YFB3303302), the National Natural Science Foundation of China (grant nos: 61977012, 62207007)
and the Central Universities Project in China for the Digital Cinema Art Theory and Technology Lab at Chongqing University (grant nos:  2021CDJYGRH011, 2020CDJSK06PT14).
\end{acks}
% \section{Acknowledgments}
% This work was supported by grants from the National Major Science and Technology Projects of China (grant no. 2022YFB3303302), the National Natural Science Foundation of China (grant nos. 61977012)
% and the Central Universities Project in China for the Digital Cinema Art Theory and Technology Lab at Chongqing University (grant nos. 2021CDJYGRH011, 2020CDJSK06PT14). 
\bibliographystyle{ACM-Reference-Format}
\bibliography{reference}%sample-base

\end{document}